\documentclass[a4paper,10pt]{article}
\usepackage{natbib}

\usepackage{subcaption}
\usepackage{amsmath}
\usepackage{amsfonts}
\usepackage{amssymb}
\usepackage{epsfig}
\usepackage{graphicx,psfrag,epsf}

\newcommand{\T}{^{\rm T}}
\newcommand{\bfX}{{\bf X}}
\newcommand{\bfD}{{\bf D}}
\newcommand{\bfU}{{\bf U}}

\newcommand{\bfM}{{\bf M}}

\newcommand{\bfW}{{\bf W}}
\newcommand{\bfP}{{\bf P}}

\newcommand{\bfV}{{\bf V}}
\newcommand{\bfE}{{\bf E}}
\newcommand{\bfQ}{{\bf Q}}

\begin{document}
%\part{Hello} 
%\chapter{Hello} 
% \title{Reduced rank matrix estimation\\ by adaptive trace norm regularization}
% \title{Adaptive trace norm low rank matrix estimation\\controlling for both thresholding and shrinkage\\using generalized SURE}
\title{Adaptive shrinkage of singular values}

\date{\today}
\author{Julie Josse\\ \emph{Agrocampus Ouest}\\and\\
Sylvain Sardy\\ \emph{Universit\'e de Gen\`eve}}

\pagestyle{myheadings}
\markright{Adaptive shrinkage of singular values}
\maketitle

\begin{abstract}

%To estimate a low rank matrix from noisy observations,
To recover a low rank structure from a noisy matrix, 
truncated singular value decomposition has been extensively used and studied.
% empirical singular values are hard thresholded and empirical singular vectors remain untouched.
Recent studies suggested that the signal can be better estimated by shrinking the singular values.
We pursue this line of research and propose a new estimator offering a continuum of thresholding and shrinking functions. %that encompasses hard and soft thresholding.
To avoid an unstable and costly cross-validation search, 
we propose new rules to select two thresholding and shrinking parameters from the data.
In particular we propose a generalized Stein unbiased risk estimation criterion that does not require
knowledge of the variance of the noise and that is computationally fast. % In addition, it estimates the rank of the matrix when the signal is detectable.
A Monte Carlo simulation reveals that our estimator outperforms the tested methods  in terms of mean squared error on both low-rank and general signal matrices across different signal to noise ratio regimes. In addition, it accurately estimates the rank of the signal when it is detectable.


\end{abstract}

%%%%%%%%%%%%%%%%%%%%%%
\section{Introduction}

In many applications such as image denoising, signal processing, collaborative filtering, it is common to model the data $\bfX$, an $N \times P$ matrix, as
\begin{equation} 
\bfX=\bfW+\bfE,
\label{eq:model}
\end{equation}
where the unknown matrix $\bfW$ is measured with i.i.d.~${\rm N}(0,\sigma^2)$ errors $\bfE$.
The matrix $\bfW$ is assumed to have low rank $R < \min(N,P)$, which means that its singular value decomposition (SVD) $\bfW=\bfP\bfD\bfQ\T$ has $R$
non-zero singular values $d_1\geq \ldots \geq d_R$. Note that model (\ref{eq:model}) is also known as bilinear model \citep{Mandel69} in analysis of variance,
as fixed factor score model \citep{deleeuw1985} or fixed effect models \citep{caussinus1986} in principal component analysis.
Such models describe well data in many sciences, such as genotype-environment data in agronomy, or relational data in social science and in biological networks,
where the variation between rows and columns is of equal interest \citep{hoff_2007_jasa}.  
   
To denoise the data, an old approach consists in performing the SVD of the matrix $\bfX=\bfU\boldsymbol{\Lambda} \bfV\T$ and
defining $\hat \bfW=\hat \bfP \hat \bfD \hat \bfQ\T$ with
$\hat \bfP=\bfU$, $\hat \bfQ=\bfV$ and keeping the first $R$ singular values while setting the others to zero.
In other words, the so-called truncated SVD  keeps the empirical directions $\bfU$ and $\bfV$, and estimates the singular values by
\begin{equation}\label{eq:hard}
\hat d_i= \lambda_i \cdot 1(i\leq R)=\lambda_i \cdot 1(\lambda_i \geq \tau),
\end{equation}
which can be parametrized either in the rank $R$ or the threshold $\tau$ (here $1()$ is the indicator function).
This estimate is also solution \citep{Eckart:1936} to 
\begin{equation}\label{eq:rank}
\min_{{\bf W} \in \mathbb{R}^{N\times P}} \frac{1}{2} \|\bfX-\bfW \|_F^2 \quad {\rm s.t.} \quad {\rm rank}(\bfW)\leq R,
\end{equation}
where $\| \bfM \|_F$ is the Frobenius norm of the matrix $\bfM$.
%Statisticians familiar with regularization techniques in regression will find the truncation (\ref{eq:hard}) and its penalty formulation (\ref{eq:rank})
%reminiscent of  best subset variable selection.
%In particular, when the regression design matrix is orthonormal, like $\bfU$ and $\bfV$ here, then the best subset has a closed form expression via hard thresholding \citep{Dono94b},
%like (\ref{eq:rank}) has a closed form expression with (\ref{eq:hard}).
The truncation (\ref{eq:hard}) and its penalty formulation (\ref{eq:rank}) are reminiscent of the hard thresholding \citep{Dono94b}
solution to best subset variable selection in regression.
The truncated SVD requires $R$ as tuning parameter, which can be selected using cross-validation \citep{Owen:cv:2009,Josse11b} or Bayesian considerations \citep{hoff_2007_jasa}. While this approach is still extensively used, recent studies \citep{Sourav:2013, Donoho:SVDHT:2013}
suggested an optimal hard threshold for singular values with better asymptotic mean squared error than thresholding at $R$ or at the bulk edge (the limit of detection).
% with better asymptotic properties than thresholding at $R$ or at the bulk edge (the limit of detection).
More precisely, \cite{Donoho:SVDHT:2013}  considered the asymptotic framework, in which the matrix size is much larger than the rank of the signal matrix to be recovered, and the signal-to-noise ratio of the low-rank piece stays constant while the matrix grows, and showed that the optimal threshold is $\left(4/\sqrt{3} \sqrt{p} \sigma\right)$ in the case of a square $(p  \times p)$ matrix and $\sigma$ known. The other thresholds for the cases of rectangular matrices and unknown $\sigma$ are also detailed in their paper.

Another popular and recent estimation strategy consists in applying a soft thresholding rule to the singular values
\begin{equation}\label{eq:soft}
\hat d_i= \lambda_i \max(1-\tau/\lambda_i,0),
\end{equation}
where any singular value smaller than the threshold $\tau$ is set to zero.
The estimate $\hat \bfW=\bfU \hat \bfD \bfV\T$ with $\hat d_i$ in (\ref{eq:soft}) is also the closed form solution \citep{Mazumder:STMiss:2010, Cai:soft:2010} to
\begin{equation} \label{eq:tracenorm}
 \min_{{\bf W} \in \mathbb{R}^{N\times P}} \frac{1}{2} \|\bfX-\bfW \|_F^2 + \tau \| \bfW \|_*,
\end{equation}
where $\| \bfW \|_*=\sum_{i=1}^{\min(N,P)} d_i$ is the trace norm of the matrix $\bfW$.
The regularization (\ref{eq:soft}) and its penalty formulation (\ref{eq:tracenorm}) are inspired by soft thresholding \citep{Dono94b} and lasso \citep{Tibs:regr:1996}. 
The tuning parameter $\tau$ is often selected by cross-validation. Recently, \citet{CandesSURE:2013} defined a Stein unbiased risk estimate (SURE) \citep{Stein:1981} to select $\tau$ more efficiently considering the noise variance $\sigma^2$ as known.
%requires an estimate of the noise variance $\sigma^2$. 

Finally, other reconstruction schemes involving nonlinear shrinkage of the singular values have been proposed in the literature \citep{Verbanck:RegPCA:2013,Nada:2013,Shabalin2013}. 
More precisely, using the same asymptotic framework as previously and asymptotic results on the distribution of the singular values and singular vectors \citep{John:asympeignull:2001,Baik:asympteig:2006, Paul:asympeig:2007}, \citet{Shabalin2013} and
\citet{OS:2014} showed that the shrinkage estimator  $\hat \bfW$ closest to $\bfW$ in term of mean squared error has the form  $\hat \bfW=\bfU \hat \bfD \bfV\T$ with, when $\sigma=1$,
\begin{equation}\label{eq:shaba}
\hat d_i = \frac{1}{\lambda_i} \sqrt{\left(\lambda_i^2 - \beta -1 \right)^2 - 4 \beta} \cdot 1\left(i\geq (1+\sqrt{\beta})\right),
\end{equation}
and $N/P \rightarrow \beta\in (0,1]$. The case with unknown $\sigma$ is also covered in their paper. 

With different asymptotics, considering that the noise variance tends to zero and $N$ and $P$ are fixed,  \citet{Verbanck:RegPCA:2013} also reached similar estimates and suggested the following heterogeneous shrinkage estimate
\begin{equation}\label{eq:twosteps}
\hat d_i = \lambda_i \left(1-\frac{\sigma^2}{\lambda_i^2}\right) \cdot 1(i\leq R).
\end{equation}
Unlike soft thresholding, the smallest singular values are more shrunk than the largest ones.
It is a two-step procedure: first select $R$, then shrink the $R$ largest singular values. % after estimating $\sigma$.
It is a compromise between hard and soft thresholding.
%The smallest singulars values are set to zero and the others are shrunk towards zero. 
%\citet{Verbanck:RegPCA:2013} showed that it often improves on hard and soft thresholding in terms of mean squared error.

In regression, \citet{Zou:adap:2006} also bridged the gap between soft and hard thresholdings by defining the adaptive lasso estimator
governed by two parameters chosen by cross-validation to control thresholding and shrinkage. To avoid expensive resampling, \citet{SardySBITE2012} selected them by minimizing a Stein unbiased estimate of the risk.
Adaptive lasso  has oracle properties and has shown good results in terms of prediction accuracy, especially using Stein unbiased risk.

%\citep{SardySBITE2012}) proposed a selection of the amount of thresholding and shrinkage by Stein unbiased risk estimation for adaptive lasso.

In this paper, we propose in Section~\ref{subsct:def} an estimator inspired by adaptive lasso to recover $\bfW$. It thresholds and shrinks the singular values in a single step using two parameters that parametrize
a continuum of  thresholding and shrinking functions.
% We recall in Section~\ref{subsct:Frob} that this estimator is solution to a weighted trace norm regularization of the least squared Frobenius norm.
We propose in Section~\ref{subsct:selection} simple though efficient strategies to select the two tuning parameters from the data, without relying on the unstable and costly cross-validation.
One approach consists in estimating the $\ell_2$-loss, the other in selecting the threshold at the detection limit, estimated empirically given the data matrix $\bfX$,
and the last one can be applied when the variance $\sigma^2$ is unknown.
%We also suggest an approach which can be applied when the variance $\sigma^2$ is unknown.
%We then show in Section~\ref{subsct:selection} how to select the  two tunning parameters from the data.
Finally, we assess the method on simulated data in Section~\ref{sec:simu} and show that it outperforms the state of the art methods in terms of mean squared error
and rank estimation.

%%%%%%%%%%%%%%%%%%%%%%%%%%%%%
\section{Adaptive trace norm}

\subsection{Definition}
\label{subsct:def}

The past evolution of regularization for reduced rank matrix estimation reveals that
the empirical singular values should not simply be thresholded (with hard thresholding),
but should also be shrunk (with soft thresholding) or more heavily shrunk, as in (\ref{eq:shaba}) and (\ref{eq:twosteps}).
% A reason for shrinking that may have been overlooked is that the empirical estimand (here the singular values of $\bfX$)
% has an upward bias \citep{denis1996, pazman1999}.
% Hence thresholding in reduced rank matrix estimation and in regression does not achieve exactly the same task
% since in regression the least squares estimate has no bias.
% For reduced rank matrix estimation, the truncation and shrinkage rule has the advantage to regulate
% not only the excess of variance but also the excess of bias of the estimation of the singular values of $\bfW$.
% In other words, for a mild regularization, the gain in mean squared error is both in variance and in bias!
Inspired by adaptive lasso \citep{Zou:adap:2006}, we propose a continuum of functions indexed by
two parameters $(\tau,\gamma)$. 
%of which we select a member $(\hat \tau,\hat \gamma)$ based on the data and some criteria discussed in Section~\ref{subsct:selection}.
The following thresholding and shrinkage function,
\begin{equation}\label{eq:adaptivelambda}
\hat d_i= \lambda_i \max\left(1-\frac{\tau^\gamma}{\lambda_i^\gamma},0\right),
\end{equation}
is defined for a positive threshold $\tau$, and encompasses soft thresholding (\ref{eq:soft}) for $\gamma=1$
and hard thresholding (\ref{eq:hard}) when $\gamma \rightarrow \infty$.
We call the associated estimator 
\begin{equation}
\label{eq:ATN}
\hat \bfW_{\tau, \gamma} = \sum_{i=1}^{\min(N,P)}  U_i \lambda_i \max\left(1-\frac{\tau^\gamma}{\lambda_i^\gamma},0\right) V_i'
\end{equation}
the adaptive trace norm estimator (ATN) with $\tau\geq 0$ and $\gamma \geq 1$.
Note that as a byproduct,
the rank of the matrix is also estimated as $\hat R= \sum_{i}^{\min(N,P)} 1(\hat d_{i} \geq 0)$.

In comparison to the hard and soft thresholding rules, the advantage of using the single and more flexible
thresholding and shrinkage function (\ref{eq:adaptivelambda}) is twofold.
First (\ref{eq:adaptivelambda}) parametrizes a rich family of functions that can more closely approach an ideal thresholding and shrinking function to recover well the structure of the underlying matrix $\bfW$, given the noise level $\sigma^2$. 
Second, the specific multiplicative factors $(1-\tau^\gamma/\lambda_i^\gamma)$ fit the rationale that
the largest singular values correspond to stable directions  and should  be shrunk mildly.
In comparison to other non linear thresholding rules, (\ref{eq:adaptivelambda}) does not rely on asymptotic derivations.
Instead it selects its parameters $(\hat \tau,\hat \gamma)$ from the data, which leads to smaller MSE in many scenarii as illustrated in Section~\ref{sec:simu}.

Our estimator is related to penalized Frobenius norm regularization.
In a matrix completion  and regression context,  \citet{Mazumder:STMiss:2010}, \citet{GaiffasWTN:2011} and \citet{CDarxivC12} proved
%For matrix completion, \citet{Mazumder:STMiss:2010} and \citet{GaiffasWTN:2011} have shown
that, for a weakly increasing weight sequence~${\boldsymbol \omega}$, the optimization problem
\begin{equation} \label{eq:weightedpenalty}
 \min_{W \in \mathbb{R}^{N\times P}} \frac{1}{2} \|\bfX-\bfW \|_F^2 + \alpha \| \bfW \|_{*,{\boldsymbol \omega}} \quad {\rm with} \quad \| \bfW \|_{*,{\boldsymbol \omega}}=\sum_{i=1}^{\min(N,P)} \omega_i d_i
\end{equation}
 has the closed form solution
$\hat \bfW=\bfU \hat \bfD \bfV\T$  with $\hat d_i= \max(\lambda_i-\alpha \omega_i,0)$.
%This is a striking result in particular when the sequence of weights is increasing.
%Indeed \citet{CDarxivC12} showed that the weighted trace is convex if and only if the sequence of weights is decreasing.
%
% Based on the thresholding and shrinkage functions discussed above (\ref{eq:hard}), (\ref{eq:soft}) and (\ref{eq:twosteps}),
% and considering the important upward biased property of the empirical singular values,
So the adaptive trace norm estimator \eqref{eq:ATN}
has weights inversely proportional to the empirical singular values, and corresponds
to $\alpha=\tau^\gamma$ and $\omega_i=1/\lambda_i^{\gamma-1}$.
%In regression, \citet{CDarxivC12} employed a related approach to estimate a matrix of linear regression coefficients to fit multiple %response vectors.
% and recommend using $\gamma=3$ based on Monte-Carlo simulations.

% Their choice of regularization parameters is driven by cross-validation for $\tau$ and Monte-Carlo simulations for $\gamma$.

%\citet{CDarxivC12} employed this approach for reduced rank regression to estimate a matrix of coefficients to fit multiple response vectors.
%Their choice of regularization parameters is driven by cross-validation and Monte-Carlo simulations.

%The simplicity of the closed form solution  contrasts with the seemingly difficult optimization problem (\ref{eq:rank}).
%Minimizing \ref{eq:tracenorm} instead of \ref{eq:rank} makes sense in the setup of matrix completion since with incomplete data since the problem is computationally hard and  \ref{eq:tracenorm} is a convex relaxation (Fazel 2002, candes 2008, mazumb 2010). 

%%%%%%%%%%%%%%%%%%%%%%%%%%%%%%%%%%%%%%%%%%
\subsection{Selection of $\tau$ and $\gamma$}
\label{subsct:selection}

%An unstable and costly selection of $\tau$ and $\gamma$ can be performed by cross-validation.
%However, in the framework of low-rank approximation matrix, it is costly. 
%Indeed, 
%the sketch of the leave-one-out cross-validation procedure is the following one. 
The parameters  $\tau$ and $\gamma$  can be estimated using cross-validation. Leave-one-out cross-validation, first consists in removing one cell $(i;j)$ of the data matrix $\bfX$. Then, for one pair $(\tau, \gamma)$, it consists in predicting its value using the estimator obtained from the dataset that excludes this cell. The value of the predicted cell is denoted $\hat X_{ij}^{-ij}$. Finally, the
prediction error is computed $(X_{ij}- \hat X_{ij}^{-ij})^2$ and the operation is repeated for all the cells in $\bfX$ 
and for each $(\tau, \gamma)$.  The pair that minimizes the error of prediction is selected. Such a procedure requires a method which provides an estimator despite the missing values. Such methods estimating the singular vectors and singular values from incomplete data exist but use computationally intensive iterative algorithms  \citep{Ilin10, Mazumder:STMiss:2010, JosseHusson12}.
This makes the cross-validation procedure difficult to use in practice even with a $K$-fold strategy. 

As an alternative, we suggest  three methods which strength is to select  $\tau$ and $\gamma$ adapting to the signal $\bfW$ and the noise level $\sigma^2$.%, without using resampling techniques.
%More precisely, we suggest three methods and investigate their finite sample properties on a Monte Carlo simulation in %Section~\ref{sec:simu}. % in terms of mean squared error and rank estimation.

\subsubsection{When $\sigma$ is known}

The first method seeks good $\ell_2$-risk. 
The mean squared error $\mbox{MSE}=\mathbb{E}||\hat \bfW-\bfW||^{2}$, or risk, of the estimator $\hat \bfW=\bfU \hat \bfD \bfV\T$ depends on the unknown $\bfW$ and cannot be computed explicitly. However, it can be estimated unbiasedly using Stein unbiased risk estimate ($\mathbb{E}(\mbox{SURE})=\mbox{MSE}$).
For estimators that  threshold and skrink singular values, SURE still has a classical form with the residual sum of squares (RSS) penalized by the divergence of the operator:
\begin{equation} 
{\rm SURE} = - NP\sigma^2 +
\mbox{RSS}+ 2 \sigma^2 {\rm div}(\hat \bfW ).
\label{eq:SUREdiv}
\end{equation}
However,  the form of the divergence is not straightforward and  \citet[Theorem 4.3]{CandesSURE:2013}  showed that it has the following closed form expression:
\begin{eqnarray*}
{\rm div}(\hat \bfW)&=&\sum_{s=1}^{\min(N,P)} \left(f'_{i}(\lambda_{i})
+ |N-P| \frac{f'_{i}(\lambda_{i})}{\lambda_{i}} \right)+ 2 \sum_{t \neq s,t=1}^{\min(N,P)} \frac{\lambda_{i} f'_{i}(\lambda_{i})}{\lambda_s^2-\lambda_t^2},
\end{eqnarray*}
where $f_{i}(\lambda_{i})$ is the thresholding and shrinking function.
The derivation of SURE requires this function to be weakly differentiable in Stein sense, which means differentiable except on a set of measure zero.
This is for instance the case for the soft-thresholding function used by \citet{CandesSURE:2013}. %(\ref{eq:soft}) that is differentiable on $\mathbb{R}$ except at $\lambda_i=\tau$.
Likewise the adaptive thresholding function (\ref{eq:adaptivelambda}) is differentiable on $\mathbb{R}$ except at $\lambda_i=\tau$ for the range of interest
$\gamma \in [1,\infty)$.
Hence SURE for adaptive trace norm  is
%For given $\tau$ and $\gamma$, the mean squared error, or risk, of the estimator $\hat \bfW=\bfU \hat \bfD \bfV\T$ using (\ref{eq:adaptivelambda})  can
%be estimated unbiasedly using Stein unbiased risk estimate.
%More precisely, we use the derivation of  \citet{CandesSURE:2013} to derive the SURE for the adaptive trace norm estimator:
\begin{equation} \label{eq:SURE}
{\rm SURE} (\tau,\gamma)= - NP\sigma^2 + \sum_{s=1}^{\min(N,P)}\lambda_s^2 \min\left(\frac{\tau^{2\gamma}}{\lambda_s^{2\gamma}}, 1\right)
+ 2 \sigma^2 {\rm div}(\hat \bfW_{\tau,\gamma}),
\end{equation}
where
\begin{eqnarray*}
{\rm div}(\hat \bfW_{\tau,\gamma})&=&\sum_{s=1}^{\min(N,P)} \left(1+(\gamma-1) \frac{\tau^\gamma}{\lambda_s^\gamma}\right) \cdot 1\left(\lambda_s \geq \tau\right)
+ |N-P| \max(1-\frac{\tau^\gamma}{\lambda_s^\gamma},0)\\
&&+ 2 \sum_{t \neq s,t=1}^{\min(N,P)} \frac{\lambda_s^2 \max(1-\frac{\tau^\gamma}{\lambda_s^\gamma},0) }{\lambda_s^2-\lambda_t^2}.
\end{eqnarray*}
A selection rule for $\tau \geq 0$ and $\gamma \geq 1$  finds the pair $(\tau, \gamma)$ that minimizes the bivariate function ${\rm SURE} (\tau,\gamma)$
in (\ref{eq:SURE}). It is not computationally costly, unlike cross-validation, but supposes the variance of the noise $\sigma^2$ as known.
%requires Gaussian noise.

The second selection method is primarily driven by a good estimation of the rank of the matrix $\bfW$. The parameter that determines the estimated rank
is the threshold $\tau$ since any empirical singular value $\lambda_i\leq \tau$ is set to zero by (\ref{eq:adaptivelambda}).
Inspired by the universal rule of \citet{Dono94b} and thresholding tests \citep{SardyANOVA13}, 
we propose to use the $(1-\alpha)$-quantile of the distribution of the largest empirical singular value $\lambda_1$ of $\bfX$ under the null hypothesis that $\bfW$ has rank zero to determine the selected threshold.
With $\alpha$ tending to zero with the sample size, null rank estimation is guaranteed with probability tending to one under the null hypothesis.
\citet{Dono94b} implicitly used level of order $\alpha=O(1/\sqrt{\log N})$ when $N=P$,
so we choose a similar level tending to zero with the maximum of $N$ and $P$.
This leads to the definition of universal threshold for reduced rank mean matrix estimation:
\begin{equation} \label{eq:tauNP}
\tau_{\max(N,P)}=\sigma F_{\Lambda_1}^{-1}\left(1-\frac{1}{\sqrt{\log(\max(N,P))}}\right),
\end{equation}
where $F_{\Lambda_1}$ is the cumulative distribution function of the largest singular value under Gaussian white noise with unit variance.
We then select the shrinkage parameter $\gamma$ by minimizing SURE (\ref{eq:SURE}) in $\gamma$ for $\tau=\tau_{\max(N,P)}$.
In practice the finite sample distribution of $\Lambda_1$ of an $N\times P$ matrix of independent and identically distributed Gaussian random variables
${\rm N}(0,1)$ is known \citep{Zanella:2009} but difficult to use. Thus, we simulate random variables from that distribution and take the appropriate quantile
to estimate $\tau_{\max(N,P)}$ in (\ref{eq:tauNP}).
Alternatively, we could use results of \citet{Shabalin2013}, who derived the asymptotic distribution of the singular values of model (\ref{eq:model}) based on results from random matrix theory \citep{John:asympeignull:2001,Baik:asympteig:2006, Paul:asympeig:2007}. 

%%%%%%%%%%%%%%%%%%%%%%%%%%%%%%%%%%%
\subsubsection{When $\sigma$ is unknown} 

Both previous methods need as an input the noise variance $\sigma^2$.
In some applications such as image denoising \citep{milanf:2013, CandesSURE:2013}, it is known or a good estimation is available. % with the median absolute deviation estimator of \citet{Dono95i}.
However, very often, this is not the case and the formula (\ref{eq:SURE}) cannot be used as such.
Inspired by generalized cross validation \citep{CravenWahba79}, we propose generalized SURE:
\begin{equation} \label{eq:GSURE}
{\rm GSURE} (\tau,\gamma)=  \frac {  \sum_{s=1}^{\min(N,P)}\lambda_s^2 \min\left(\frac{\tau^{2\gamma}}{\lambda_s^{2\gamma}}, 1\right)    }
{(1-{\rm div}(\hat \bfW_{\tau,\gamma})/(NP))^2}.
\end{equation}
% GSURE corresponds to the residual sum of squares penalized by an unbiased estimation of the degrees of freedom. 
Using a first order Taylor expansion $1/(1-\epsilon)^2$ of \eqref{eq:GSURE}, we get that $\mbox{GSURE}\approx\mbox{RSS}\left(1+2\ {\rm div}(\hat \bfW_{\tau,\gamma})/(NP)\right)$;
then considering the estimate of variance $\hat \sigma^2=\mbox{RSS}/(NP)$, 
one sees how GSURE approximates SURE \eqref{eq:SUREdiv}. 

The GSURE criterion has the great advantage of not requiring any input value and can be applied straightforwardly
to select both tuning parameters.
% Alternatively, leave-several-out cross validation could be implemented to estimate the risk as a function of $(\tau, \gamma)$
% by removing random units in the $\bfX$ matrix and predicting them.

%%%%%%%%%%%%%%%%%%%%
\section{Simulations}
\label{sec:simu}

\subsection{Gaussian setting with large $N$ and $P$}

We compare the adaptive trace norm estimator to existing ones by reproducing the simulation of \citet{CandesSURE:2013}.
Here, matrices of size $200\times 500$  are generated according to model \eqref{eq:model} with four signal-to-noise ratios
SNR $\in \{0.5, 1, 2, 4\}$ (calculated as one over $\sigma \sqrt{NP}$) and two values for the rank $R \in \{10, 100\}$.
For each combination, 50 datasets are generated.
%The criteria of interest are mean squared error and rank that we estimate by taking the median of the 50 values calculated.
We consider five estimators:
\begin{itemize}
 \item Truncated SVD (TSVD). We use the common one with the true rank $R$ for (\ref{eq:hard}) as well as the ones proposed by \citet{Donoho:SVDHT:2013} with asymptotic MSE optimal choices of hard threshold $\tau=\lambda_{*}(\frac{N}{P})\sqrt{P}\sigma$ when $\sigma$ is known,
 and $\tau=w(\frac{N}{P}) {\rm median}(\lambda_i)$ when $\sigma$ is unknown.
 The values for the coefficients $\lambda_{*}(\frac{N}{P})$ and  $w(\frac{N}{P})$ are given in their Tables 1 and 4.
 \item Optimal shrinkage (OS) of \citet{Shabalin2013} and \citet{OS:2014} when $\sigma$ is known as well as when $\sigma$ is unknown as defined in Section 7 of \citet{OS:2014}.
 \item Singular value soft thresholding (SVST) with $\tau$ selected to minimize SURE \citep{CandesSURE:2013} for (\ref{eq:soft}), knowing $\sigma$.
 \item the 2-step estimator with the true rank $R$ \citep{Verbanck:RegPCA:2013} for (\ref{eq:twosteps}).
 \item Adaptive trace norm (ATN) with three selections of the two  parameters indexing a family of shrinkers.
 With $\sigma$ known: SURE (\ref{eq:SURE}), and SURE as a function of $\gamma$ only ($\tau$ is set to the universal threshold (\ref{eq:tauNP})).
 With $\sigma$ unknown: GSURE (\ref{eq:GSURE}).
\end{itemize}
We report in Table~\ref{simu_candes} and~\ref{simu_candes*} the
estimated mean squared error between the fitted matrix $\hat \bfW$ and the true signal $\bfW$, and the estimated rank (the number of singular values that are not set to zero).
We also include the lower bound on worst-case MSE for any matrix denoiser as given in \citet{Donoho:minimaxsvst:2014} which can be used as a baseline. The standard deviations of the MSEs are very small for all the estimators and vary from the order of $10^{-5}$ for high SNR to $10^{-3}$ for small SNR. Thus, the MSEs can be directly analysed to compare the estimators. We indicate the standard deviations for the rank.

Table~\ref{simu_candes} reports the performance with no oracle information, while Table~\ref{simu_candes*} is when some parameters are known, either the true rank or the true noise variance.
Comparing the two Tables allows to assess the performance loss by having to estimate all parameters, like in most real life applications.

% Only two estimators are fully automatic and are reported in the first two columns: the proposed adaptive trace norm estimator with GSURE and the truncated SVD (\citep{Donoho:SVDHT:2013}).
% Columns with  the sign $*$  mean that the estimator assumes knowledge of either the true rank or the true variance $\sigma^2$, for benchmark.

\begin{table}
\begin{center}
\scriptsize 
\begin{tabular}{rr||r|r|r|r}
\multicolumn{1}{r}{$R$}	&	\multicolumn{1}{c||}{SNR}	&
\multicolumn{1}{c|}{ATN} & \multicolumn{1}{c|}{TSVD} &
\multicolumn{1}{c|}{OS}  & Lower\\ 
        &               &  \multicolumn{1}{c|}{{\bf GSURE}} & \multicolumn{1}{c|}{${\boldsymbol \tau}$} && bound 
           \\ \hline
        {\bf MSE} &&&& &\\
10&4&\textbf{0.004}& \textbf{0.004} & \textbf{0.004}& 0.0012 \\
100&4&\textbf{0.037}&0.409 & 0.335 &0.0106\\
10&2&\textbf{0.017}&\textbf{0.017} &  \textbf{0.017} &0.0024\\
100&2&{\bf 0.142}&0.755 & 0.606 &0.0212\\
10&1&\textbf{0.067} & 0.072 & \textbf{0.067}&0.0048\\
100&1&\textbf{0.454}&1.000 & 0.892&0.0424\\
10&0.5&\textbf{0.254}& 0.321 & \textbf{0.250} & 0.0097\\
100&0.5&\textbf{0.978}&1.000 & 0.994& 0.0845\\ \hline
{\bf Rank}&&&&\\
10&4& 11 (1.8) & \textbf{10} (0.0)&  \textbf{10} (0.0)\\
100&4& \textbf{102} (1.7)& 49 (1.2) &  78 (0.7)\\
10&2&  11 (1.4) & \textbf{10} (0.0)&  \textbf{10} (0.0)\\
100&2& \textbf{112} (2.8)& 20 (1.6)&  48 (1.3)\\
10&1& 11 (1.3)& \textbf{10} (0.0)& \textbf{10} (0.0)     \\
100&1& \textbf{140} (4.3)& 0 (0.0)& 16 (1.6)\\
10&0.5& 15 (1.6)& \textbf{10} (0.0)&  \textbf{10} (0.0)\\
100&0.5& \textbf{14} (7.6)& 0 (0.0)& 2 (1.2)\\ \hline
\end{tabular}
\caption{Monte Carlo results in terms of mean squared errors (top) and rank estimation with its standard deviation (bottom). $R$~is the true rank (10 or 100) and SNR is the signal-to-noise ratio.
Three fully automatically estimators are considered: adaptive trace norm (ATN) based on GSURE (\ref{eq:GSURE}), truncated SVD (TSVD) using $\tau=\omega(0.4) {\rm median}(\lambda_i)$ \citep{Donoho:SVDHT:2013}
and optimal shrinkage (OS) using the estimation for the noise variance \citep{OS:2014}.  The lower bound of \citet{Donoho:minimaxsvst:2014} is indicated.
Sample size is $N = 200$ individuals and number of variables is $P = 500$. Results correspond to the mean over the 50 simulations.
Best results linewise are indicated in {\bf bold}.
%Values reported with a number of digits of the order permitted by the Monte Carlo simulation.
\label{simu_candes}}
\end{center}
 \end{table}
 \normalsize

\begin{table}
\begin{center}
\scriptsize 
\begin{tabular}{rr||r|r|r|r|r|r|r}
\multicolumn{1}{r}{$R$}	&	\multicolumn{1}{c||}{SNR}	&
\multicolumn{2}{c|}{ATN} & 
\multicolumn{2}{c|}{TSVD} & \multicolumn{1}{c|}{OS} & \multicolumn{1}{c|}{SVST} & \multicolumn{1}{c}{2-steps}  \\ \hline
        &               &  
        \multicolumn{1}{c}{${\rm SURE}$}  & \multicolumn{1}{c|}{${\rm universal}$} &        
        \multicolumn{1}{c}{$R$} & \multicolumn{1}{c|}{$\tau$} & \multicolumn{1}{c|}{} & \multicolumn{1}{c|}{${\rm SURE}$}     &  \multicolumn{1}{c}{} 
           \\ \hline
        {\bf MSE} &&&&&&& \\
10&4 & \textbf{0.004}&\textbf{0.004}&
\textbf{0.004} & \textbf{0.004} &  \textbf{0.004} &0.008&\textbf{0.004}\\
100&4&\textbf{0.037}&\textbf{0.037}&
0.038 &0.038 &  \textbf{0.037}   & 0.045&\textbf{0.037}\\
10&2&\textbf{0.017}&\textbf{0.017}&
\textbf{0.017} & \textbf{0.017} &   \textbf{0.017}     & 0.033&\textbf{0.017}\\
100&2 &{\bf 0.142}&0.147&
0.152 & 0.158  &   0.146    & 0.156&\textbf{0.141}\\
10&1& \textbf{0.067}&\textbf{0.067} &
0.072 & 0.072 &  \textbf{0.067}     & 0.116&\textbf{0.067}\\
100&1&\textbf{0.448}&0.623&
0.733 &0.856 &  0.600    & \textbf{0.448}&0.491\\
10&0.5&0.253& 0.251&
0.321 & 0.321 &  \textbf{0.250}   & 0.353&0.257\\
100&0.5&\textbf{0.852}&0.957&
3.164 &1.000 &  0.961    &\textbf{0.852}&1.477\\\hline
{\bf Rank}&&&&&&&& \\
10&4&   11 (1.6)& \textbf{10} (0.0) &   &  \textbf{10} & \textbf{10}   & 65  (2.3)&   \\
100&4& 103 (1.9)&  \textbf{100} (0.0)&  & \textbf{100} & \textbf{100}    & 193 (0.8) &  \\
10&2&    11 (1.0)&  \textbf{10} (0.1)&  & \textbf{10}  & \textbf{10}    & 63  (2.2)&  \\
100&2 &114 (2.2)& \textbf{100} (0.0)&  & \textbf{100}  & \textbf{100}    & 181  (1.1)&  \\
10&1&   11 (1.2)&  \textbf{10} (0.0)&   & \textbf{10}  & \textbf{10}  & 59 (1.7)  &    \\
100&1 &154 (1.8)&  \textbf{65} (0.8)&  & 38 (0.6)&  64  (0.7) & 154 (1.2) &  \\
10&0.5 &  15 (1.6)& \textbf{10} (0.0) &   & \textbf{10} &  \textbf{10}    & 51 (2.7)&   \\
100&0.5 & \textbf{87} (3.3)&   16 (0.8)& &   0  &  15  (0.8)  & 86  (2.6)&  \\ \hline
\end{tabular}
\caption{Same setting as for the Monte Carlo simulations of Table~\ref{simu_candes}.
The same estimators are considered, except that oracle quantities, either the true rank $R$ or the true variance $\sigma^2$, are used.
Considered estimators are: ATN with two selection rules, SURE and universal, for the two parameters, assuming $\sigma$ is known;
TSVD knowing true rank $R$ and $\tau=\lambda_*(0.4)\sqrt{500}\sigma$ \citep{Donoho:SVDHT:2013};
Optimal shrinkage (OS) with $\sigma$ known \citep{OS:2014};
singular value soft thresholding (SVST) with $\sigma$ is known \citep{CandesSURE:2013};
two-steps knowing true rank $R$ \citep{Verbanck:RegPCA:2013}.
% Best results overall linewise are indicated in {\bf bold}.
\label{simu_candes*}}
\end{center}
 \end{table}
 \normalsize

\begin{figure} 
    \begin{center}
    \includegraphics[scale=0.43, angle=-90]{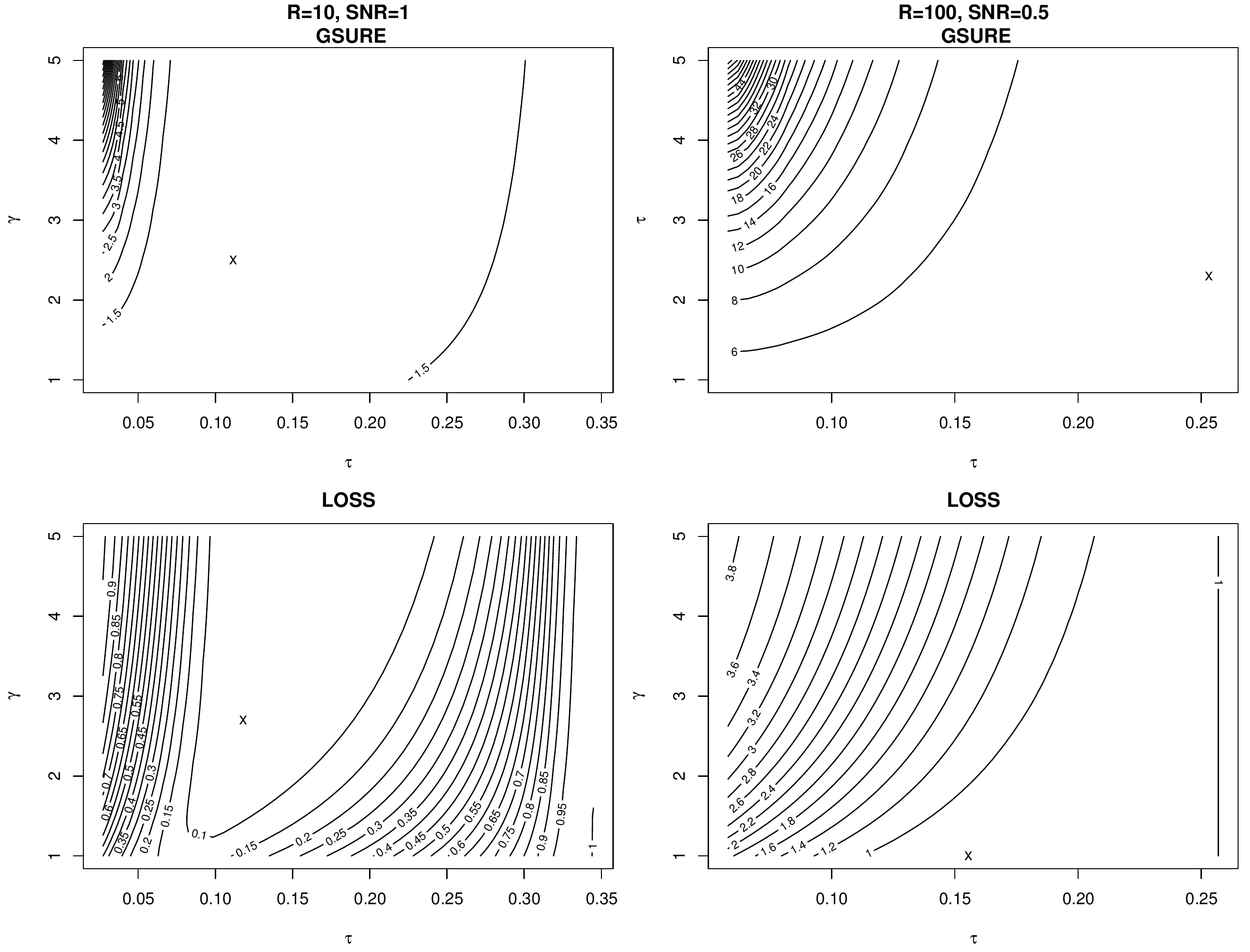}
 \caption{Top: Generalized Stein unbiased risk estimate (GSURE); Bottom: corresponding true $\ell_2$-loss. Both are plotted as a function of $(\tau,\gamma)$ for data generated
 as in Table~\ref{simu_candes}. Left: true rank $R=10$ and signal to noise ratio ${\rm SNR}=1$; Right: $R=100$ and ${\rm SNR}=0.5$. The cross 'x' points to the minimum of the bivariate
 curve. Comparing columnwise, we see good fit between the location of the minima, especially on the left.  \label{fig:tau_gamma}}
    \end{center} 
  \end{figure}

Looking at Table~\ref{simu_candes}, we see that  the proposed adaptive trace norm estimator
performs remarkably well, owing to its flexibility (with two parameters) and to a good selection of the appropriate model with GSURE. It can even outperform oracle estimators (in Table~\ref{simu_candes*}) that are governed by a single parameter. Here GSURE results are very similar to its corresponding SURE results, showing that \eqref{eq:GSURE} is a good approximation to \eqref{eq:SUREdiv} thanks to a large value for $NP$.
Figure~\ref{fig:tau_gamma}  illustrates the striking ability of GSURE to approximate the true loss in different regimes. % by an appropriate selection  $(\hat \tau, \hat \gamma)$.
On the top, the estimated risk of ATN GSURE is represented as a function of $(\tau,\gamma)$ for $R$=10 and SNR=1 on the left (row 6 of Table \ref{simu_candes}) and for $R$=100 and SNR=0.5 (last row of Table \ref{simu_candes}) on the right.
On the bottom, the true loss function is plotted as a function of $\tau$ and $\gamma$. The estimated values (top) and the optimal choice (bottom) located by a cross
are close.

\begin{figure} 
    \begin{center}
    \includegraphics[scale=0.43, angle=-90]{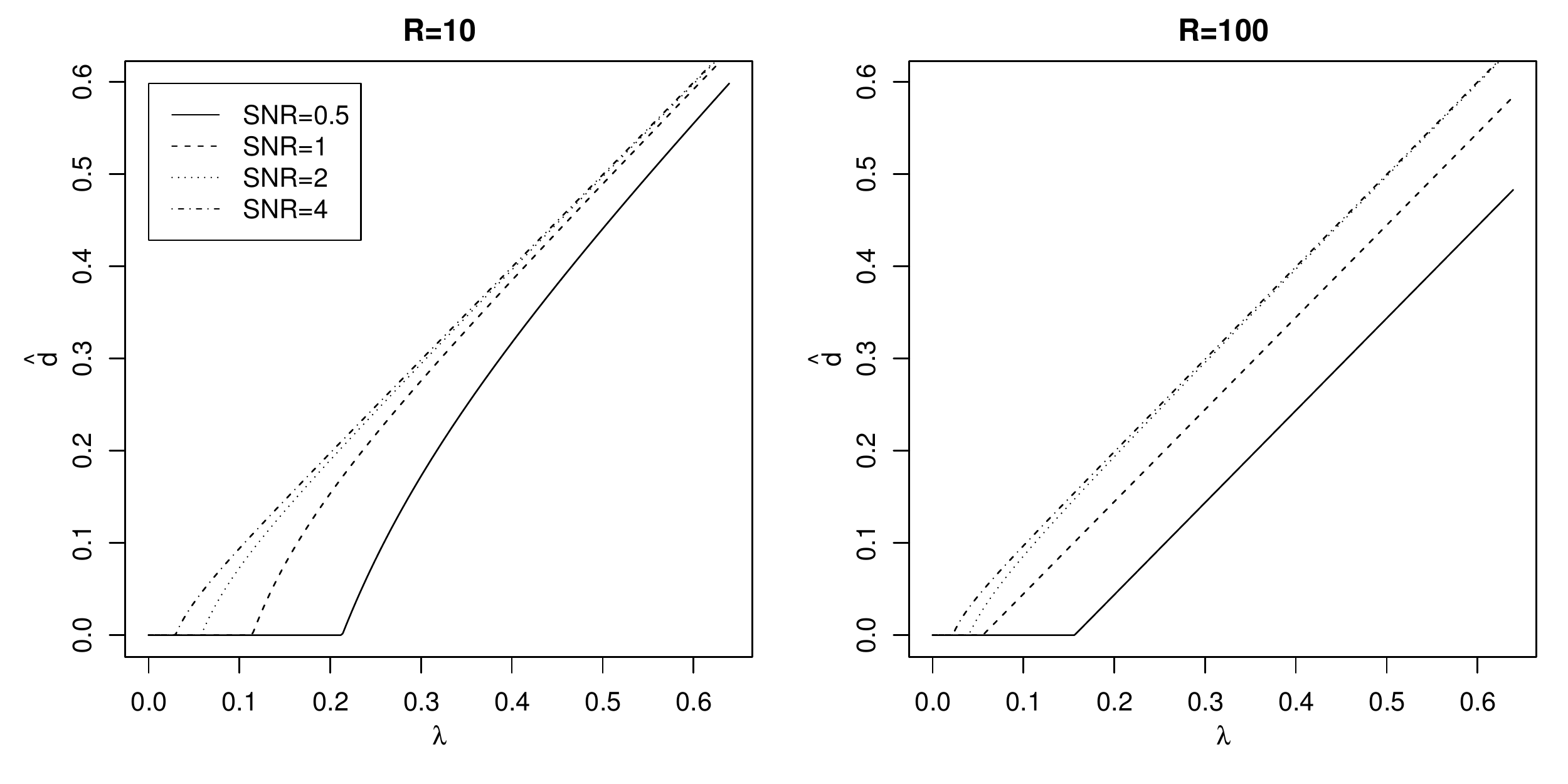}
 \caption{Typical thresholding function selected by ATN with SURE for ranks $R=10$ (left) and $R=100$ (right), and for four values of SNR. \label{fig:shapeshrink}}
    \end{center} 
  \end{figure}

Looking at Table~\ref{simu_candes*}, we see that when the SNR is high (equal to 4 and 2), the ATN, 2-step, TSVD (with $R$ and $\tau$) and OS estimators give results in term of MSE of the same order of magnitude and clearly outperform the SVST approach.
When the SNR decreases, the TSVD, and the 2-step and OS methods to a lower extend, collapse;  
this is when the SVST provides better results especially for the difficult setting when the rank $R=100$.
%Owing to its two regularization parameters well chosen, the adaptive trace norm method ``adapts" its behavior to the data and outperforms all the other methods.
% More precisely, the adaptive approach with the tuning parameter selected using SURE$^*$ (column 6) is as good as the TSVT (versions $R^*$ and $\tau^*$) when the signal is very strong,
% as good as the 2-step method, and finally as good as the SVST approach when the signal is very noisy (rows 6 and 8).
The good behavior of the 2-step approach in many situations highlights the fact that it is often a good strategy to apply a different amount of shrinkage to each singular value.  These simulations provide good insights into the regimes for which each estimator is well suited: low noise regime for the TSVD, moderate noise regime for the 2-step,
and high noise for the SVST. But, if one is interested in a single estimator, regardless of the unknown underlying structure,
then ATN becomes the method of choice.
Figure \ref{fig:shapeshrink} illustrates the adaptation of ATN to various SNR and rank by
representing a typical shrinking and thresholding function selecting $(\tau,\gamma)$ with SURE. 
As expected when SNR is 0.5, ATN is close to soft thresholding, whereas when SNR is 4, it is close to hard thresholding. 
% This is because the threshold $\tau$ increases when the SNR decreases.
% Concerning $\gamma$, it is equal to 1 for the difficult situations when the rank is large and the data are very noisy meaning that the ATN estimator only resorted to soft thresholding to recover the signal.  

As far as rank estimation is concerned,  even if it is not the primary objective, ATN gives a very good estimation of the rank and SVST considerably over-estimates. This phenomenon is also known in the setup of regression where the lasso tends to 
select too many variables \citep{Zou:adap:2006,Zhang:2008}. Note that ATN using the universal threshold (column 2 of  Table~\ref{simu_candes*}) was designed for estimating the rank and indeed provided a very good estimate. 

Finally, the results of TSVD and OS (oracle or not) when $R=100$ %reported in Tables~\ref{simu_candes} and~\ref{simu_candes*}.
are not as competitive as when $R=10$, because these methods are based on asymptotics and assume low rank compared to matrix size.
%which is not satisfied here when $R=100$, $N=200$ and $P=500$.
In addition, the MSEs are very similar across both Tables for $R=10$ whereas there are differences for $R=100$ which indicates that the estimation of $\sigma$ encounters difficulties. 

The case SNR=0.5 and $R$=100 in Table~\ref{simu_candes*} is also worth a comment.
Here, the data are so noisy that part of the signal is indistinguishable from the noise:
only 16  singular values are greater than the ones that would be obtained under the null hypothesis that the rank of the $N\times P$ matrix $\bfW$ is zero. 
Nevertheless, ATN with SURE estimates on average a matrix of rank 87, which is quite remarkable.
In this situation, the same amount of shrinkage  is applied to all the singular values with soft thresholding (see that the selected $\hat \gamma=1$ on Figure~\ref{fig:shapeshrink} in this situation)
leading to the smallest MSE.

\subsection{Non-Gaussian noise}

\begin{figure} 
    \begin{center}
    \includegraphics[width=\textwidth]{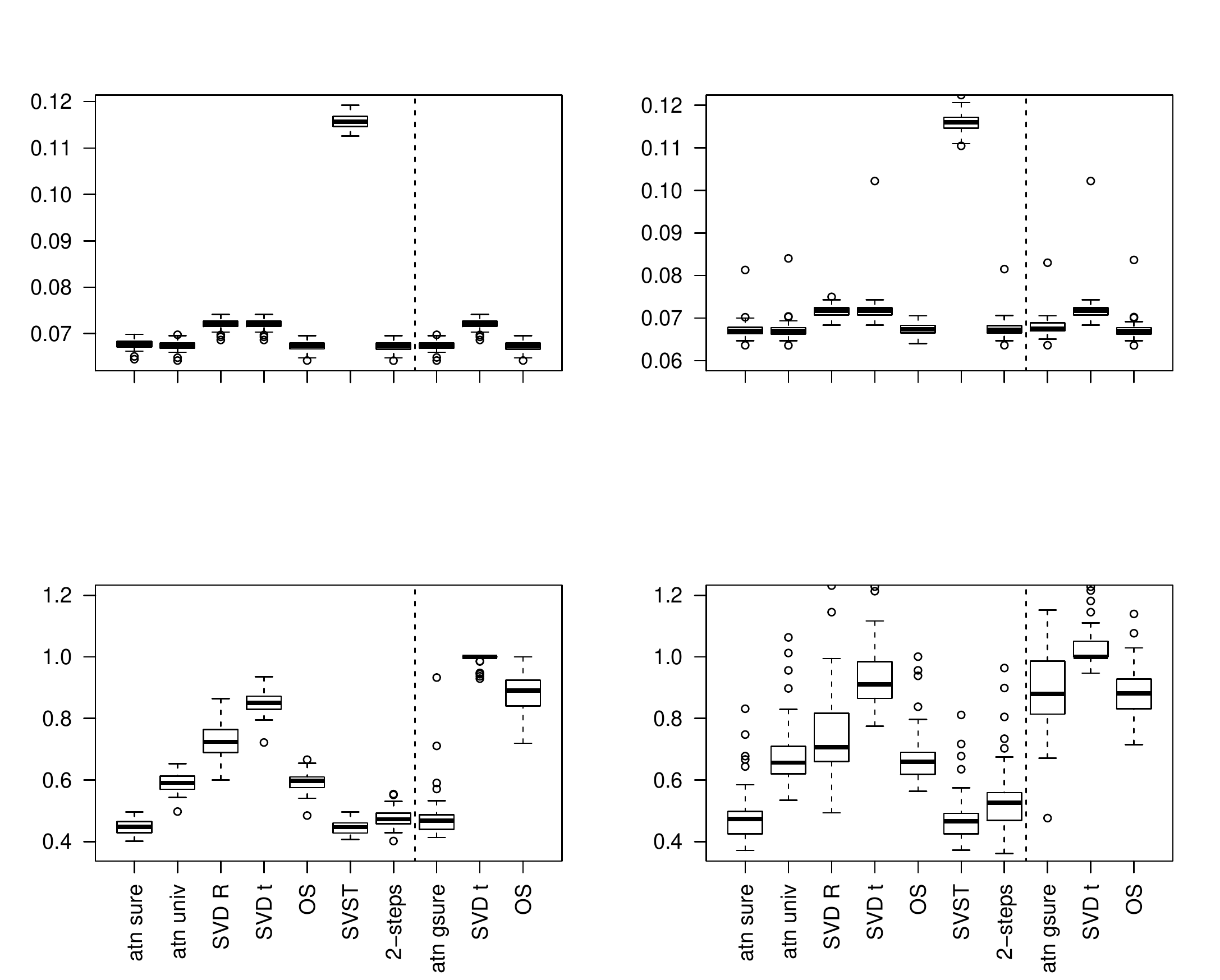}
 \caption{MSE boxplots for $R=10$ and SNR=1. Top: $N=200$, $P=500$; bottom: $N=20$, $P=50$. Left: Gaussian; right: Student. 
 \label{fig:ptitgauss_stud}}
    \end{center} 
  \end{figure}

The methods considered above are all based on the assumption of Gaussian noise.
We now assess their sensitivity to Student noise with 5 degrees of freedom, based on the same simulations as for Table~\ref{simu_candes}.
We also considered a more difficult situation with a matrix size divided by 10, that is $N=20$ and $P=50$.
Figure~\ref{fig:ptitgauss_stud} points to two noticeable consequences: the boxplots are more variable with Student, yet centered around the same median,
except for ATN with GSURE that sees its efficiency drop when both $N$ and $P$ are small (bottom right).
% 
% Surprisingly, all the estimators are quite robust to this heavy tail noise: the MSEs are centered around the same values,
% the only difference being  a slight increase in their variability as illustrated on the top of Figure \ref{fig:ptitgauss_stud}.
% %However, this is only to a small extent and does not modify any comments on the behavior of the estimators.
% These  good results can be explained by the fact that the noise remains symmetric.
% We managed to give a rough time to the estimators and particularly to the ATN with GSURE either by using a Student noise with 3 degrees of freedom or when generating data both with Student noise and of small size.
% Figure \ref{fig:ptitgauss_stud} on the bottom shows, in this latter case, that the variance of all the estimators increases but ATN with GSURE really collapses. 

%%%%%%%%%%%%%%%%%%%%%%%%%%%%%%%%%%%%%%%%%%%%%%%%%%
\subsection{Simulations based on a real small data set}
% 
% In this section, our aim is to extend the simulation study to provide more information about the efficacity of the proposed methods by generating data of different size
% varying the rank and the SNR.

We consider here a realistic simulation based on a wine dataset with $N=21$ wines described by $P=30$ sensory descriptors (the data are available in the R package FactoMineR \citep{Le:Josse:Husson:2008:JSSOBK:v25i01}). We used the fitted rank-$R$ matrix
as the true signal matrix, and then added Gaussian noise to perform a Monte Carlo simulation. 
Note that in practice, it makes sense to center the data before using the estimators since the values are shrunk toward zero. 
\begin{figure}
\begin{center}
\includegraphics[scale=0.5]{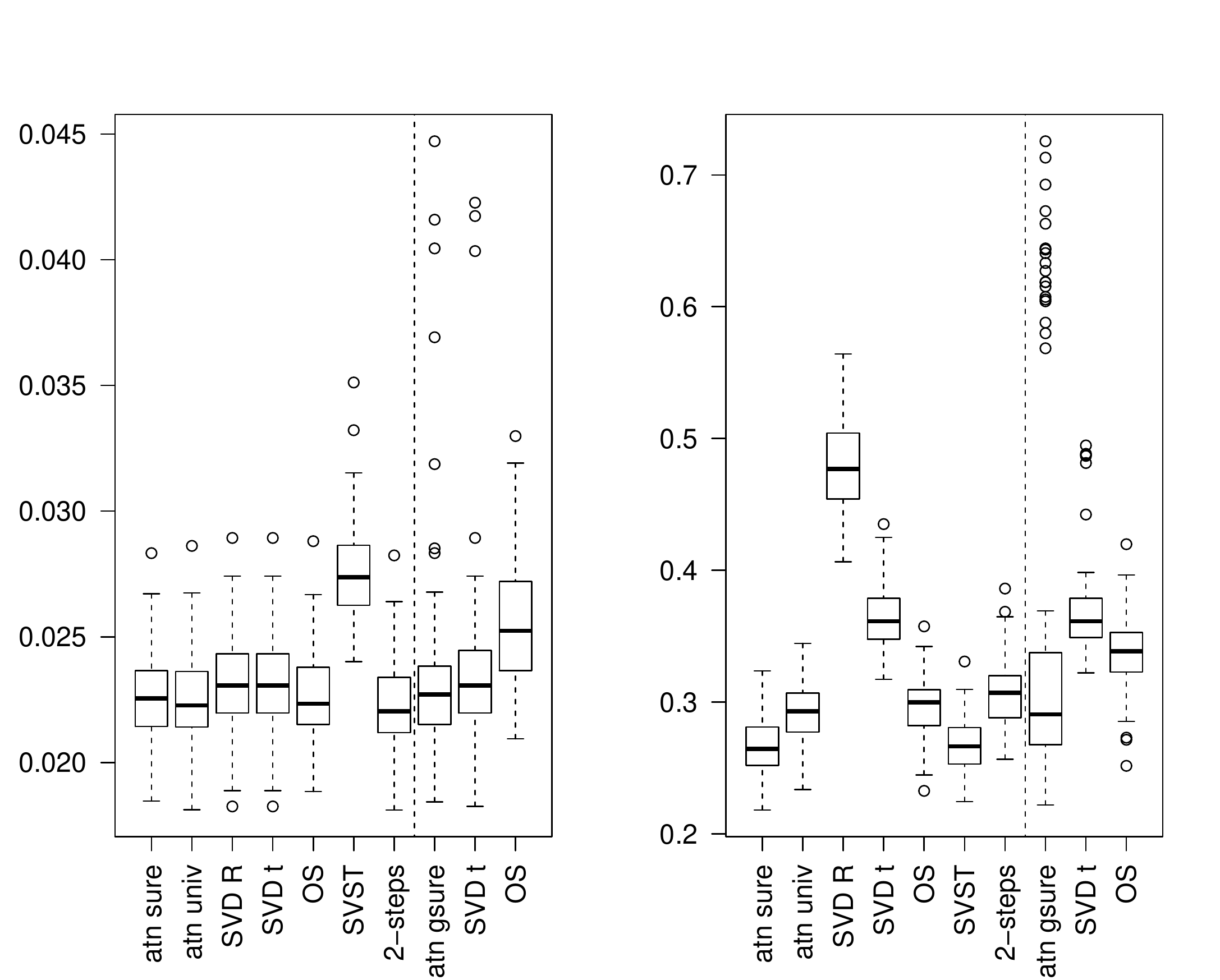}
\caption{Distribution of the MSE for the wine dataset simulations for $R=8$.} 
\label{fig:ptit}
\end{center}
\end{figure}
On this small sample case, we found the same trends as observed previously. As illustrated in Figure~\ref{fig:ptit} for a case with two levels of noise and $R$=8, the estimators often manage to improve on the usual truncated SVD, 
the SVST fits well for high noise regime and poor otherwise, and ATN remains very powerful. %. %Note that the optimal estimators are surprisingly accurate in this %setting although,
%as mentioned previously, their performances deteriorate with high ranks especially for unknown noise variance.
Note that ATN with universal $\tau$ has results similar to those of the optimal shrinker (OS), which provides an empirical interpretation of the OS estimator and highlights the capability of ATN to find the optimal way to shrink the singular values.
Finally,  GSURE, although  still the best method among the blind estimators (the last 3 estimators on the graphics) in term of median MSE, is more variable.  

%followed ... and generated for the former a student distribution with 5 degrees of freedom $t_5$ and for the latter $t_5+exp(1)-1$ %where exp is the exponential distribution with expectation 1.
%This results are representative of many simulations we performed. However, we will the 
%and right skewed noise

%%%%%%%%%%%%%%%%%%%%
\section{Conclusions}

Recovering a reduced rank matrix  from noisy data is a hot topic that has aroused the scientific community for a few years, as testified by the abundant recent literature on the subject. 
The adaptive trace norm estimator combines the strength of the hard, soft and the two-step procedures by means of a shrinking and a thresholding parameter indexing a family of shrinkers. The method adapts to the data which ensures good estimation of 
both low rank and general signal matrices, whatever the regime encountered in practice.
The tuning parameters are estimated without using computationally intensive resampling methods thanks to the SURE and GSURE formulae.
The latter version has the great advantage of not requiring knowledge of the noise variance. 
In addition, the rank is also estimated accurately, especially with the universal version designed for it. Our method outperforms the competitors on simulations and thus can be recommended to users. 

We showed that Student noise affected the results, %all the estimators appear quite robust to the violation of the Gaussian noise hypothesis. However, 
but other corruptions such as strong outliers alter the performances of the estimators to a greater extent.
To tackle this issue, a natural extension could be to derive robust estimators using for instance the robust Huber loss function $\rho$ instead of the Frobenius norm in (\ref{eq:weightedpenalty}). It leads
to an estimator solution to
\begin{equation} \label{eq:robustATN}
\min_{{\bf W} \in \mathbb{R}^{N\times P}} \|\bfX-\bfW \|_\rho + \alpha \| \bfW \|_{*,{\boldsymbol \omega}} \quad {\rm with} \quad \| \bfW \|_{*,{\boldsymbol \omega}}=\sum_{i=1}^{\min(N,P)} \omega_i d_i,
\end{equation}
where $\| H \|_\rho=\sum_{i=1}^N \sum_{j=1}^P \rho(h_{ij})$ with
$$
\rho(h)= \left \{
\begin{array}{ll}
h^2/2, & |h|\leq  \tilde \alpha \\
\tilde \alpha |h|-\tilde \alpha^2/2, & |h|> \tilde \alpha
\end{array}
\right .
$$
is the Huber loss with cutpoint $\tilde \alpha$ \citep{Huberbook}. Extending the results of \citet{STB01}, we could rewrite  (\ref{eq:robustATN}) as
$$
\min_{{\bf W},{\bf R} \in \mathbb{R}^{N\times P}, } \|\bfX-\bfW - {\bf R} \|_F^2 + \alpha \| \bfW \|_{*,{\boldsymbol \omega}} + \tilde \alpha \| {\bf R} \|_1 \quad {\rm with} \quad
\| {\bf R} \|_1 = \sum_{i=1}^N \sum_{j=1}^P |R_{ij}|.
$$
It would allow to solve the problem by block coordinate relaxation and could be used as an alternative to the robust estimator of \citet{robustPCA09}.

Two others extensions should be considered. First,  assessing our estimator in a missing data framework as an alternative to 
iterative soft thresholding algorithms \citep{Mazumder:STMiss:2010}. Second, assessing our estimator to denoise inner product matrices or  covariance matrices as in \citep{Ledoit:nonlinshrink:2012}.
Finally, we can mention the work of \cite{hoff_2013} who suggested a Bayesian treatment of Tucker decomposition methods to analyze arrays datasets. 
% His method also regularizes least squares via the eigenvalues.
%Thus,  it also leads to denoise the data and to get a better estimator in term of mean squares error.
%To better fit data, he pointed to hierarchical priors where the values of the hyper-parameters could be learnt from the data %with an empirical Bayesian approach. His proposal follows
%essentially the same goal as our method.
To better fit data, he pointed to hierarchical priors learning the values of the hyper-parameters from the data with an empirical Bayesian approach, which follows essentially the same goal as our method.

The results are reproducible with the {\tt R} code provided by the first author on her webpage. 

\section*{Acknowledgment}

The authors are grateful for the helpful comments of the reviewers and editors.
J.J. is supported by an AgreenSkills fellowship of the European Union Marie-Curie FP7 COFUND People Programme. S.S. is supported by the Swiss National Science Foundation.
This work started while both authors were visiting Stanford University and the authors would like to thank the Department of Statistics for hosting them and for its stimulating seminars.

\bibliographystyle{plainnat}
\bibliography{article}

\end{document}